\documentstyle[bo99,epsfig]{article}

\title{Obscured Active Galactic Nuclei}
\author{R. Maiolino}
\affil{Osservatorio Astrofisico di Arcetri,
 Firenze, Italy}

\begin{document}

\maketitle

\begin{abstract}
The properties of the absorption in type 2, narrow line AGNs are reviewed by
focusing on the X-ray indicators. I discuss the properties of the cold
absorbing medium (the putative torus)
and of the reprocessed components, as well as their implications for the
unified model. The relation between optical classification and
X-ray absorption is examined.
The case of ``fossil'' AGNs, whose
type 2 classification is not due to absorption effects, is also discussed.
Although this review is mainly focused on nearby Seyfert 2 galaxies,
I also shortly discuss
the effects of absorption at higher luminosities and higher
redshift and the implications for the X-ray background.
\keywords{galaxies: active -- galaxies: Seyfert -- galaxies: ISM --
X-rays: galaxies}
\end{abstract}

\section{Introduction.}

Several observational data indicate that type 2, narrow line Seyfert nuclei
suffer significant obscuration along our line of sight. The unified model
(Antonucci 1993) ascribes this obscuration to a gaseous parsec-scale
circumnuclear torus. According to this model, broad line
 Seyfert 1 (Sy1) and narrow line Seyfert 2
(Sy2) galaxies
would be identical physical objects, while the orientation of the line
of sight with respect to the torus is responsible for the obscuration
of the Broad Line Region (BLR) and of the nuclear engine (X--UV source) in
Sy2s.

In this review I show how X-ray observations, especially at high
energies ($>$2 keV), provide a wealth of information which constrain the
properties of the absorbing medium in Seyfert 2s, i.e. the putative 
circumnuclear gaseous torus. I also discuss the implications of these
X-ray studies on related issues, such as the fuelling of active
nuclei and the X-ray background.
Due to the limited space, I do not discuss warm absorbers, although
evidence for this component is found in some obscured Seyferts
(eg. Komossa \& Fink 1997), neither I discuss
the soft excess that
characterizes most Sy2s (Maiolino et al. 1998, Wilson \& Elvis 1997).


\section{X-ray absorption as a diagnostic of the torus.}

X-ray observations have probably provided the most direct test of the unified
model. Indeed, the X-ray spectrum of many Sy2s is characterized by a
powerlaw similar to that observed in Sy1s and a photoelectric cutoff
due to absorbing, cold gas along our line of sight. Since the absorbing column
in Sy2s is generally larger than $\rm 10^{22}cm^{-2}$, these studies were
mostly obtained with satellites sensitive in the high energy band above
$\sim 1 keV$ such as Ginga, ASCA and BeppoSAX.

If the absorbing column is larger than
$\sim 10^{24} cm^{-2}$, i.e. the medium is thick to Compton scattering, then
the transmitted component is completely suppressed below 10 keV and the
spectrum observed in the 2--10 keV band is dominated by reprocessed components.
More specifically, in this case the hard X-ray spectrum is characterized
by a flat Compton reflection component, ascribed to the inner surface of
the torus, and/or a steeper component due to a ionized, warm scattering
medium (eg. Matt et al. 1997). If the absorbing medium 
has a column in the range $10^{24}-10^{25}cm^{-2}$,
then the transmitted component can be
still observed in the 10--300 keV band as an emission excess (eg. Done et al.
1996); for larger absorbing columns even the 10--300 keV band is dominated
by the reflection component. The cold reflector also produces a prominent
fluorescence iron line at 6.4 keV (EW$\sim$1--2 keV with respect to the
reflected continuum). Yet, Netzer et al. (1998) pointed out
 that also the Narrow Line
Clouds can contribute significantly to the observed 6.4 keV line, provided
that their column density is large enough. Instead,
the warm scattering medium emits lines of He-- and H--like iron at 6.7 and
7 keV, which can reach large equivalent widths as well
 (Matt et al. 1996).
Until a few years ago only a couple of Compton thick sources were known. This
is because the reflection efficiency of both the cold and warm mirrors is
low ($\sim 10^{-2}-10^{-3}$) and, therefore, this class of objects is more
difficult to detect. This highlights one of the major problems in past
hard X-ray surveys, as discussed in the following section.

\subsection{The distribution of N$_H$.}

Past hard X-ray surveys of Sy2s were strongly biased in favor of X-ray bright
sources (according to all-sky surveys), which tend to be the least absorbed
ones. More recently, X-ray observations have probed X-ray weaker Sy2s,
partly removing the selection against heavily obscured objects (eg.
Turner et al. 1997). We (Salvati et al. 1997, Maiolino et al. 1998) used
BeppoSAX to observe an [OIII]-selected sample of previously unobserved
(X-ray weak) Sy2s. These observations discovered
 a large fraction of Compton thick objects, a
result which confirms the bias against heavily obscured systems affecting
previous surveys. However,
none of these surveys is actually complete or free from biases
and, therefore, none of them is suitable (if taken separately) to derive
the distribution  of absorbing column densities among Sy2s. To accomplish
this goal, we merged
all the available hard X-ray observations of Sy2s and extracted a complete
subsample limited in {\it intrinsic} (i.e. unabsorbed) luminosity as
inferred from the [OIII] narrow emission line (Risaliti et al. 1999).
This subsample is composed of 45 objects and the corresponding N$_H$
distribution, shown in Fig.~1, can be considered the best estimate of the
true distribution that can be obtained
with the available data. The most interesting result is that
this distribution is significantly shifted toward large columns with respect
to past estimates: most ($\sim$75\%) of the Sy2s are heavily obscured
($\rm N_H > 10^{23} cm^{-2}$) and about half are Compton thick. The N$_H$
does not appear to correlate with the (intrinsic) luminosity, at variance
with early results.
This N$_H$ distribution has various implications some of which will be
discussed in the following.

\begin{figure}[!h]
\centerline{
\begin{minipage}[c]{0.5\linewidth}
\psfig{file=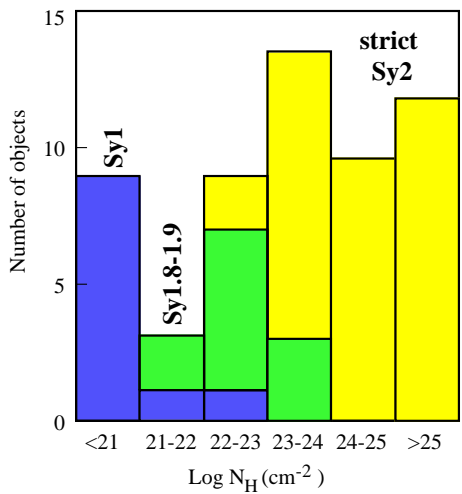,width=6cm}
\end{minipage}
\begin{minipage}[c]{0.5\linewidth}
\caption{
The distribution of absorbing column densities among Seyfert galaxies.
(from Risaliti et al. 1999, with modifications).}
\end{minipage}
}
\end{figure}

\subsection{X-ray absorption as a probe of the parsec scale torus.}

The large fraction of Compton thick sources constrains the size of the
obscuring medium. Indeed, assuming an axisymmetric geometry, the mass of the
gas responsible for the obscuration is about $\rm M_{gas}\approx f~N_H R^2$,
where R is the external radius of the torus and f is the covering factor
which, for the Compton thick medium, must be of the order of 0.5. As
discussed in Risaliti et al. (1999), the requirement that the gas mass does not
exceed the dynamical mass constrains the external radius of the torus to
$\rm R < 10~pc$. Enhanced metallicity could help to relax this constraint.
Metallicities a few times solar have actually been inferred for the BLR and
in the warm absorber of several Sy1s, but an order of magnitude higher
metallicities would be required to relax the mass constraints enough to
match the size of the torus of a few
100 pc proposed by some models (sect. 2.3).

A parsec-scale size of the obscuring torus is also supported by the
variability of N$_H$. Indeed, in several Seyfert galaxies the photoelectric
cutoff is observed to change significantly on time scales of a few years
(Malizia et al. 1997). Fig.~2 shows the specific case of NGC7582 (from a
compilation of results in Turner et al. 2000, Xue et al. 1999 and references
therein, where the same model was adopted to fit the data).
Although in some cases differences in N$_H$ measured at different
epochs might be ascribed to the different satellite used, in some cases
variation of N$_H$ are observed even with the same satellite.
Finally, I recall that a parsec-scale component of the obscuring torus has
been directly observed in radio VLBI images, both in continuum (free-free)
and H$_2$O maser emission, in a few active nuclei (eg. Gallimore et al.
1997, Greenhill \& Gwinn 1997).

\begin{figure}[!h]
\centerline{
\begin{minipage}[c]{0.6\linewidth}
\psfig{file=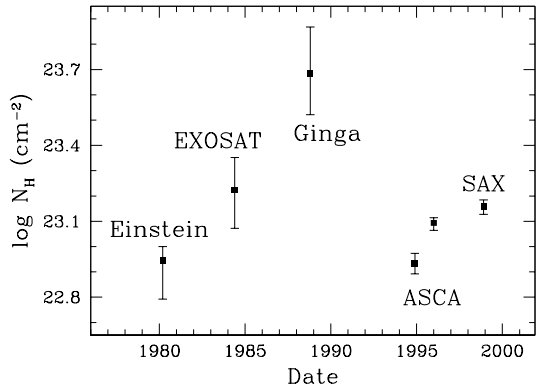,width=7cm}
\end{minipage}
\begin{minipage}[c]{0.4\linewidth}
\caption{
Variation of the absorbing N$_H$ in the Seyfert 2 galaxy NGC7582.}
\end{minipage}
}
\end{figure}

\subsection{X-ray absorption as a probe of the 100 pc scale torus.}

X-ray studies have also provided evidence for obscuring medium on larger
scales ($\sim$100 pc). Indeed, samples of AGNs selected in the soft X-rays
($\le 2$ keV) are characterized by a shortage of edge-on galaxies
(Lawrence \& Elvis 1982, McLeod \& Rieke 1995, Simcoe et al. 1997). This
effect is ascribed to circumnuclear material, associated to the galactic
gaseous disk, which obscures the AGN when the host galaxy is oriented edge-on.
Only a moderate absorbing column ($\rm N_H\ge 10^{22} cm^{-2}$) is required to
suppress significantly the soft X-ray emission.

A 100 pc scale component of the obscuring torus is also required to 
fit the broad IR spectral energy distribution (Granato et al. 1997)
and it is also directly observed in the HST images of some objects
(eg. Ford et al. 1997, Malkan et al. 1998).

\subsection{Intermediate type Seyferts.}

Very likely, the obscuring torus has both a Compton thick parsec-scale
component and a larger 100 pc-scale component which is
coplanar with the galactic disk and characterized by lower N$_H$
(which does not violates the mass constraints). The internal torus obscures
completely the BLR along the intercepting lines of sight (Sy2 case), while the
obscuration due to the external torus (in edge-on systems) is moderate and
the Seyfert nucleus  might show weak broad lines and, therefore, might
be classified
as intermediate type 1.8--1.9. Indeed, intermediate type Sys are more
commonly found in edge-on galaxies (Maiolino \& Rieke 1995). This model is
also supported by the distribution of absorbing column of intermediate
Seyferts, which is significantly shifted towards lower values of
N$_H$ than strict Sy2s (Risaliti et al. 1999, Fig.~1).
Although, this model might apply to many
intermediate type Seyferts, there are Seyfert galaxies
 whose intermediate properties are not
to ascribe to moderate obscuration but to photoionization and
variability effects or to more complex forms of absorption
(Goodrich et al. 1995, Komossa \& Fink 1997).

\subsection{Dual absorbers.}

A simple, absorbed power law does not provide an adequate description of
the observed X-ray spectrum for some of the
Compton thin Sy2s. In these cases there
is evidence for a second photoelectric cutoff at higher energies due to
a medium that only partially covers the X-ray source
(Malaguti et al. 1999, Hayashi et al. 1996, Weaver et al.  1994, Vignali
et al. 1998). While the photoelectric cutoff at low energies is very likely
due to the torus, the partial covering observed at higher energies can only
be obtained if the second absorbing medium is close to the X-ray source
and has a similar size. A most likely candidate is the BLR.
Generally, the column density of the partial covering medium is
of the order of $\rm \sim 10^{23} cm^{-2}$, in agreement with estimates for
the N$_H$ of broad line clouds. The covering factor of the partial
absorber is generally found to be larger than $\sim$ 30\%, that is significantly
higher than the covering factor of the broad line clouds which, based
on the broad lines equivalent width and on the absence of the Ly-edge cutoff
in the UV spectrum of QSOs, is expected to be about
 10\%. However, I do not
consider this a major caveat since the dual absorbers discovered so far
are not representative of the true distribution of the partial covering
absorption systems, but only sample the tail with high covering factor.
Partial covering systems with a covering factor of 10\% probably remained
undetected in the
past observations. I expect Chandra and XMM to discover a large
number of dual absorbers with low partial covering.

Finally, thanks to the extended spectral coverage of BeppoSAX some cases of
dual absorber with partial covering characterized by an absorbing
column as high as $\rm N_H \sim 10^{24} cm^{-2}$ are being found
(eg. Turner et al. 2000). Such a high N$_H$ is still consistent with that
expected for the broad line clouds. Indeed, the estimated column of
$\rm 10^{22}-10^{23} cm^{-2}$ for the broad line clouds is only a lower
limit that is required to produce the low ionization broad emission lines
(MgII, FeII, etc...).

\section{X-ray absorption and galaxy morphology.}

Although the column density of the cold obscuring material does not depend
on the properties of the AGN (eg. its luminosity) it does depend on the
properties of the host galaxy. Indeed, the absorbing N$_H$ strongly
correlates with the presence of a stellar bar in the host galaxy (Maiolino et
al. 1999). In particular, while non-barred Sy2s are characterized by an
average $\rm log~N_H \sim 22~(cm^{-2})$, most of the strongly
barred Sy2 galaxies are Compton thick. This finding indicates that stellar
bars are effective in driving gas into the nuclear region to obscure the
AGN. This result is in line with other studies, as reviewed by
Sakamoto (1999), which indicate that bars are effective in driving
gas into the nuclear region, though these other studies are
not specifically focused on AGNs. We
might speculate that, more generally, non-axisymmetric potentials
(eg. distorted morphologies and galaxy interactions) drive gas into the
nuclear region. Although there are some observational indications in this
direction, a
systemic study, similar to that on barred systems, has not been performed
yet. This will be possible with the new Chandra and XMM
data.

\section{The relation between optical and X-ray absorption.}

If the obscuring torus has the same gas-to-dust ratio as in the Galactic
ISM, and the dust is characterized a Galactic extinction curve, then
the nuclear region of Sy2s should suffer a visual extinction that is related to
the gaseous column density by the formula $\rm A_V=5\times 10^{-22} N_H
(cm^{-2})$. In general this is not the case: $\rm A_V$ is lower than
expected from the N$_H$ measured in the X-rays. This was first pointed out
by Maccacaro et al. (1982). A visual extinction lower than that expected
from the N$_H$ measured in the X-rays is also required to fit the IR
spectrum of AGNs (Granato et al. 1997). We have collected a sample of Seyferts
which both show X-ray (cold) absorption and whose optical or IR broad lines
are not completely suppressed. The ratios between the broad lines provide
information on the dust reddening towards the nucleus; however, the broad
emission lines must be used with much care, since the extreme conditions
of the broad line clouds can affect the intrinsic line ratios through
radiative transport effects.
By assuming the standard extinction curve we can estimate the
visual extinction.
The resulting distribution for the $\rm A_V/N_H$ ratio, relative to the
Galactic standard value, is shown in Fig.~3.
Most of the AGNs in our sample are characterized by a
deficit of dust absorption with
respect to what expected from the N$_H$ measured in the X-rays, in
agreement with early claims. At higher, quasar-like luminosities there are
even more extreme examples of this effect:
objects that, although absorbed in the X-rays, do not show significant dust
absorption in the optical and appear as type 1, broad line AGNs
have been recently
discovered in hard X-ray and radio surveys (Sambruna et al. 1999,
Akiyama et al. 2000, Reeves et al. 1997). 
Puzzling enough, the early Chandra surveys presented to date have found
only a few type 1 QSOs absorbed in the hard X-rays; this issue will be shortly
discussed in Sect.~7.

The origin of the reduced $\rm A_V/N_H$ ratio is not clear.
An obvious explanation is that the
dust-to-gas ratio is much lower than Galactic
or that in the inner part of the obscuring torus the dust is sublimated by
the strong UV radiation field. However, if the dust content in the absorbing
medium is significantly reduced, especially at the inner face, then
most of the UV ionizing photons are absorbed by the atomic gas.
This should create a huge HII region, which would emit
strong ($\sim$ narrow) hydrogen lines corresponding to a large
covering factor, i.e. much brighter than the emission lines from the NLR
(see also Netzer \& Laor, 1993). Also, a simple shortage of dust grains with
respect to the gas mass would not explain other peculiar properties of the
dust in AGNs, such as the absence of the silicate absorption feature in the
mid-IR spectra of most Sy2s (Clavel et al. 2000) and the absence of the
carbon dip in the UV spectra of some reddened Sy1s.

Another interesting
possibility is that the dust extinction curve is much flatter than the
standard Galactic. The high density of the gas
 in the circumnuclear region of AGNs
is likely to favor the growth of large grains (probably through coagulation)
which, in turn, should flatten
the extinction curve and make it featureless.
This effect is directly observed in the dense clouds of
our Galaxy (Draine 1995).
Within the context of the optical versus X-ray absorption, the effect
of a flat extinction curve (due to grain coagulation)
is twofold: 1) given the same dust mass, the
effective visual extinction is lower, and 2) the broad lines ratio gives a
deceiving (low) measure of the extinction. A more thorough discussion
of the whole issue is given in Maiolino et al. (2000b).

\begin{figure}[!h]
\centerline{
\begin{minipage}[c]{0.5\linewidth}
\psfig{file=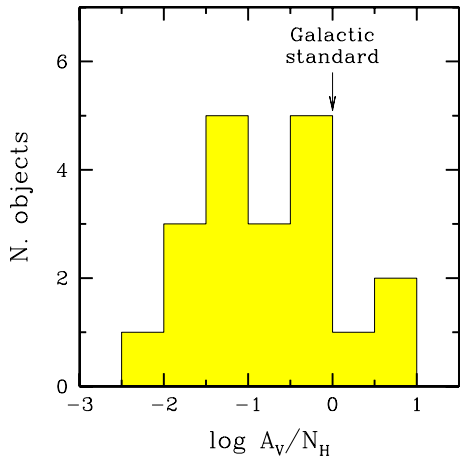,width=6cm}
\end{minipage}
\begin{minipage}[c]{0.5\linewidth}
\caption{
Distribution of the $\rm A_V/N_H$ ratio, relative to the Galactic standard
value, for a sample of absorbed AGNs.}
\end{minipage}
}
\end{figure}

\section{Piercing completely hidden AGNs.}

So far I have discussed X-ray absorption in AGNs which were discovered and
classified in the optical. However, an increasing number of obscured 
powerful AGNs has
been discovered by means of hard X-ray observations in galaxies which are
optically classified as starburst or LINER. Probably, in these objects the
obscuring medium hides also the NLR. Alternatively, the nuclear ionizing
source might be completely embedded and obscured in all directions. The
most spectacular case is the nearby (4 Mpc) edge-on galaxy NGC4945, which
hosts one of the brightest AGNs at 100 keV (Done et al. 1996)
The powerful X-ray nucleus is obscured by a column of
$\rm \sim 5 \times 10^{24} cm^{-2}$ along our line of sight. However, it
seems that the nucleus is heavily obscured in all directions. Indeed, optical
to mid-IR observations were unable to detect any indication of the AGN
activity: at these wavelengths
the central region is characterized by a starburst with a spectacular
superwind cavity where LINER-like lines are produced
(Maiolino et al. 2000a, Marconi et al. 2000).
A case quite similar to NGC4945, but at much higher luminosities, is
NGC6240. The optical spectrum of this strongly interacting
system shows only
weak LINER-like emission lines and the mid-IR properties  are similar to
starburst galaxies (Lutz et al. 1999). However, hard X-ray observations
have detected the presence of a heavily obscured AGN whose intrinsic
luminosity is in the QSO range (Vignati et al. 1999). Other luminous IR
galaxies, optically classified as starburst or LINER, might also host
completely hidden AGNs (Risaliti et al. 2000).

Some of the luminous hard X-ray sources recently discovered by deep
Chandra surveys have optical and near-IR counterparts that show little or no
evidence for AGN activity (eg. Mushotzky et al. 2000, Fiore et al. 2000).
Possibly, many of these objects are the analogous
at higher redshift of the local, completely hidden AGNs discussed above.

\section{Fossil Active Nuclei.}

Not all of the narrow line, type 2 Seyfert galaxies are associated to
absorption along our line of sight. A fraction of the Seyfert galaxies have
shown a strong decrease of their X-ray flux, on time scales of
several years, that is not associated to an increased absorption but to
an intrinsic drop of their activity (eg. Bassani et al. 1999b). One of the
best studied cases is NGC2992 (Weaver et al. 1996). The fraction of these
``fossil'' AGNs is about 10\% (Maiolino et al. in prep.). However, this is
actually a lower limit to their real fraction since the identification of
this class of objects requires X-ray observations at different epochs and, in
particular, the source had to be in a high, bright state during the
observations performed by the early, low sensitivity missions. These
conditions are met only for a limited number of Seyferts. Obviously, the
conservation of the number of active nuclei requires fossil AGNs to revive
after a certain period, as it is actually observed (eg. Gilli et al. 2000a).

Shortly after that an AGN turns off, the BLR fades and the nucleus appears
as a type 2 Seyfert whose X-ray emission is dominated by the cold reflection
component due to the circumnuclear torus (Guainazzi et al. 1998).
Although the AGN appears as a Compton thick Sy2, these features are
{\it not} due to obscuration. About 10 years later also the echo from the
torus should fade. If the nucleus remains in a quiescent state for an even
longer period then also the gas in NLR clouds
 can recombine. Yet, the high ionization regions
of the clouds should recombine much faster and, therefore, the observed
narrow line spectrum should be characterized by low ionization lines similar
to LINERs (Eracleous et al. 1995). In this phase the X-ray emission
should be dominated by the {\it non-variable} warm reflection component with,
possibly a highly ionized Fe line at 6.7--7 keV, as indeed observed in
several LINERs. Some of the LINER nuclei might actually be
fossil AGNs.

\section{Implications for the X-ray background.}

Obscured AGNs are thought to be a key ingredient of the hard X-ray
background (XRB, Setti \& Woltjer 1991, Comastri et al. 1995). This is
discussed in detail in this volume by other authors (eg. Comastri).
However, I wish to emphasize that some of the issues discussed in this review
might have implications for our understanding of the origin of the hard XRB.

The distribution of N$_H$ represents the main set of free parameters in the
XRB synthesis models. The N$_H$ distribution presented in sect.~2.1 can be
used to freeze this set of parameters, under the assumption that the
distribution does not evolve with redshift. Detailed models that take
into account this constraint are presented in Gilli et al. (1999, 2000b):
although the shape and power of the XRB is well reproduced,
the observed number counts in the hard X-rays seem to require an evolution
of the obscured AGNs faster than for the unobscured population.
This result is also
related to the finding that non--axisymmetric morphologies increase the
obscuration of the
active nuclei (sect.~3). Indeed, the latter effect should have
an impact on the XRB, since galaxies at high redshift are characterized by
a higher rate of distorted/irregular morphologies.

The deficiency of dust absorption, with respect to the X-ray absorption,
especially at high luminosities,
implies a possible mismatch between optical and X-ray
classification of the sources contributing to the hard X-ray background.
In particular, some of the type 2 QSOs, which are expected to make
most of the hard XRB, could be optically ``masked'' as
type 1 QSOs and already present in optical surveys.
As mentioned in Sect.~4, the early Chandra surveys have identified only
a few objects of this class, at variance with what found in radio and past
hard X-surveys (Sect.~4). Possibly the discrepancy is due to the
sensitivity of Chandra which peaks in the soft X-ray band
and, therefore, probably biases the surveys in favor of little absorbed
sources. To properly tackle this issue we should wait for the identification of
a larger number of Chandra sources (especially at fainter fluxes, where
the fraction of absorbed sources is higher) and the results of the
surveys that are being performed with XMM, whose sensitivity is much more
uniform up to $\sim$ 7 keV.

\begin{acknowledgements}
Many of the new results presented in this paper were obtained in collaboration
with R. Gilli, A. Marconi, G. Risaliti and M. Salvati.
This work was partially supported by the Italian Space Agency (ASI)
through the grant ARS-99-15 and by the Italian Ministry for University and
Research (MURST) through the grant Cofin-98-02-32.
\end{acknowledgements}

\end{document}